\documentclass[12pt]{article}

\usepackage{pdfpages}
\usepackage[utf8]{inputenc}
\usepackage{authblk}
\usepackage{graphicx}
\usepackage{amssymb}
\usepackage{amsmath,amsthm}
\usepackage{url}
\usepackage{color}
\usepackage{natbib}
\usepackage[defaultfam,light,tabular,lining]{montserrat} 

\usepackage{setspace}
\usepackage{caption}
\usepackage{parskip}
\usepackage[utf8]{inputenc}

\usepackage[top=0.8in, bottom=1.6in, left=1in, right=1in]{geometry}
\title{Industrial complexity and the evolution of formal employment in developing cities}

\author[1,2,4,*]{\small Neave O’Clery}
\author[3]{Juan Chaparro}
\author[4]{Andres Gomez-Lievano}
\author[4,5]{Eduardo Lora}
\affil[1]{Centre for Advanced Spatial Analysis, University College London}
\affil[2]{Mathematical Institute, University of Oxford}
\affil[3]{School of Finance, Economics and Government, Universidad EAFIT}
\affil[4]{Harvard Growth Lab, Harvard University}
\affil[5]{Research in Spatial Economics, Universidad EAFIT}
\affil[*]{Corr. Author: n.oclery@ucl.ac.uk}

\date{\small \today}

\begin{document}

\maketitle

\vspace{4cm}

\footnotesize{The authors acknowledge Matte Hartog, Alfredo Guerra and José Ramón Morales for their assistance with data collection and cleaning, and Juan Pablo Chauvin, Ricardo Hausmann and Carmen Pages-Serra for helpful comments and discussion. We acknowledge financial support from Fundación Mario Santo Domingo. This article was completed with support from the the PEAK Urban Program, supported by UKRI’s Global Challenge Research Fund, Grant Ref: ES/P011055/1.}

\clearpage

\vspace*{3cm}
\begin{center}
   \Large Industrial complexity and the evolution of formal employment in developing cities 
\end{center}

\vspace{1cm}

\begin{abstract}

What drives formal employment creation in developing cities? We find that larger cities, home to an abundant set of complex industries, employ a larger share of their working age population in formal jobs. We propose a hypothesis to explain this pattern, arguing that it is the organised nature of formal firms, whereby workers with complementary skills are coordinated in teams, that enables larger cities to create more formal employment. From this perspective, the growth of formal employment is dependent on the ability of a city to build on existing skills to enter new complex industries. To test our hypothesis, we construct a variable which captures the skill-proximity of cities’ current industrial base to new complex industries, termed ’complexity potential’. Our main result is that complexity potential is robustly associated with subsequent growth of the formal employment rate in Colombian cities.

\end{abstract}

\clearpage

\section{Introduction}

A key open question in economic development is the factors and dynamics that govern the growth of the formal sector. Here, we focus on the local labour market as a key driver of the dynamics of formal employment creation in sophisticated sectors. Specifically, we propose a framework in which formal firms are composed of, in essence, organised teams of skilled and complementary workers which are found more abundantly in larger cities. Formal firms engage in complex activities which depend on access to new and complementary skills for growth. This framework leads us to our main hypothesis, namely that the growth of the formal employment rate, via the growth of existing formal firms or the entry of new formal firms, is dependent on the potential of a city to enter new complex industries by building on the skills embedded in the local labour market. 

A multitude of alternative definitions delineate formal and informal activities \citep{MAChen2012WIEGOworkingpaper1, la_porta_informality_2014, kanbur2017informality}. While we do not focus primarily on the transition from informal to formal jobs, the distinction between formal and informal jobs and firms is important for our work. In general, jobs in which individuals work in formal firms remain limited in emerging and developing countries, and 'formal employment' (defined by the ILO as employment in incorporated enterprises) employs only 28 percent of the working age population in developing countries\footnote{Average of 35 countries with comparable data for 2015 from ILO. See \url{https://ilostat.ilo.org/data/}.} Our conceptual framework is based on the distinction or difference in character between formal and informal firms. Specifically, within our framework, a formal firm is a registered firm which contributes to social security, while an organised firm is one that combines the skills of several complementary workers in order to generate output. We argue that formal firms tend to be also organised firms. Informal firms, on the other hand, tend to be run by individuals or families, and operate in low complexity activities requiring few highly complementary skills. The use of the word ``organised'' to refer to formal firms is not uncommon, and is, for example, the official terminology in India's National Commission for Enterprises in the Unorganised Sector \citep{nceus2007}. Our work can be seen as providing theoretical and empirical support for such use. 

\begin{figure}[t!]
    \centering
  \includegraphics[width=0.8\textwidth]{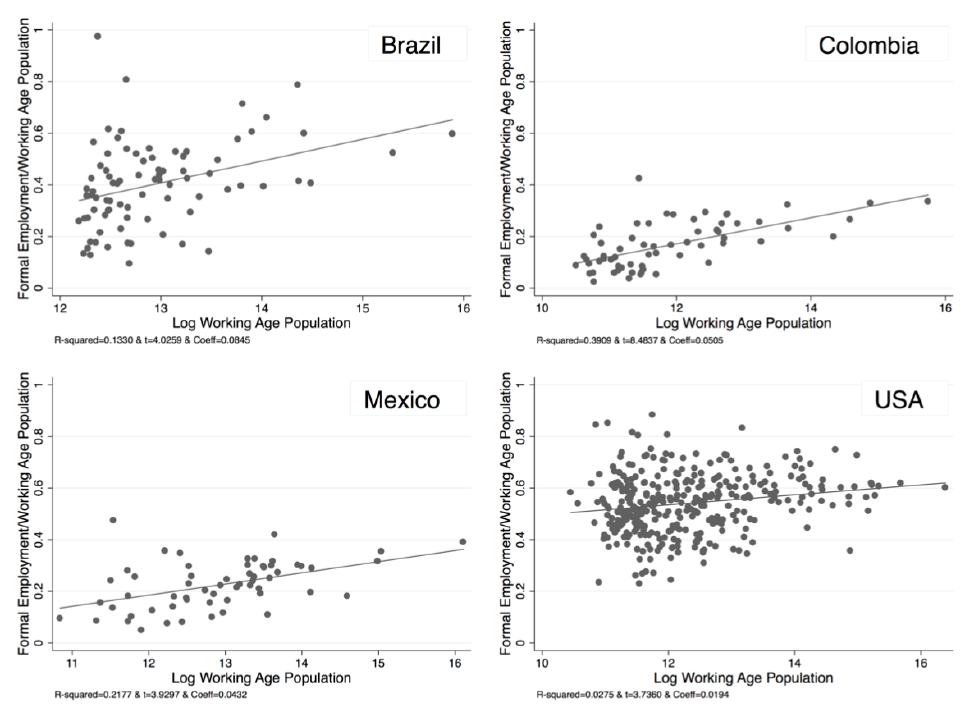}
  \caption{The share of the working age population employed by formal firms is higher in larger cities across Brazil, Colombia, Mexico and the USA.}
  \label{fig_2}
\end{figure}

This study is based on the idea that the combinations of productive capabilities that are required by complex industrial activities are more easily found in larger cities with a more diverse labour force. As a result, more organised employment is created in larger cities. Our proxy for organised employment, the formal employment rate (formal employment as a share of the working age population), is consistently higher in larger cities as shown in Figure \ref{fig_2} for Brazil, Colombia, Mexico and the United States. As the figure suggests (and the corresponding statistical tests underneath corroborate), formal firms employ relatively more people in cities with larger working age populations (similar patterns have been uncovered by, e.g., \citealp{gomez-lievano_explaining_2017, balland2020complex, davis_comparative_2020}). This idea is also consistent with a variety of studies that show that larger cities are home to a larger overall diversity of occupations \citep{bettencourt2014professional}, and firms with a more diverse range of occupations \citep{tian2021division}.  

\begin{figure}[t!]
    \centering
  \includegraphics[width=0.8\textwidth]{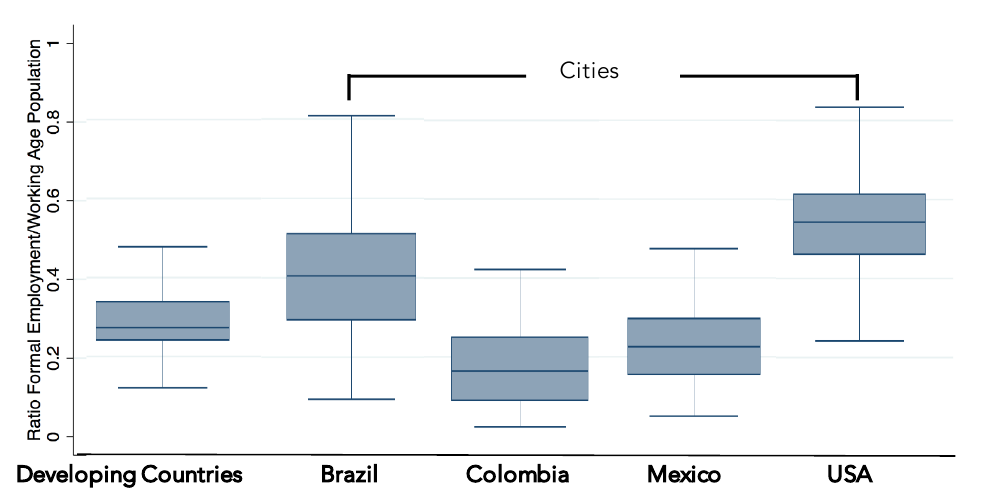}
     \caption{Box plots for the distribution of formal occupation rates in a set of developing countries, and cities in Brazil, Colombia, Mexico and the USA. All data for cities is for 2013 except Brazil which is 2010. The boxes contain the observations between the 25th and the 75th percentiles. We observe a larger variance in formality rates across cities within countries than across countries, suggesting that the study of the determinants of formal employment across cities is an important area of research.}
  \label{fig_1}
\end{figure}

Not only is the formal employment rate higher in larger cities, but the variance is also  much higher across cities than across countries, as seen in Figure \ref{fig_1}. We can interpret this figure as suggesting that the key drivers of labour formality (such as the availability of skills) are to be found at a local, and more specifically at the city level, and not at a national level. While some studies have shown that regional differences in institutions affect labour informality \citep{jonasson2012government, almeida2009enforcement} and the size of the informal sector \citep{chaudhuri2006size}, none have focused on differences across cities. We posit that there is a significant gap in our current understanding of the mechanisms through which cities make formal employment grow. 

We propose a novel mechanism to explain why formal employment in developing countries with abundant supplies of labour is higher in larger cities. We argue that the availability in the local labour force of skill combinations needed in sophisticated activities determines the potential for growth of the formal employment rate. This hypothesis is consistent, in turn, with the association between informality and low productivity \citep{kanbur2017informality, la_porta_informality_2014}, since we expect firms combining more diverse and complementary skills to be more productive. In order to test our hypothesis, we construct a city-wide variable called ``complexity potential'', which captures the skill-relatedness between a city's existing industry basket and the set of complex industries that the city is currently missing (see Section~\ref{sec_indcomplskrel}). We implement this variable using formal employment data for cities in Colombia and show that it predicts the growth of the formal employment rate for Colombian cities between 2008 and 2016 controlling for relevant factors. 

Our main contribution is to the literature on economic development, and namely the determinants of the growth of formal employment. Specifically, we go beyond the traditional focus on the capabilities of governments (both national and local) to enforce the labour and tax codes \citep{perry2007informality, ulyssea_regulation_2010}, or the more recent focus on the strategic decisions of workers and firms in response to the costs and benefits of becoming formal \citep{narita2020, meghir_wages_2015}, and focus on the local availability of skills as the main restriction facing firms to become more sophisticated and create more formal employment. In other words, we focus not on the transition of informal firms to formality, but on the processes by which formal employment grows in complex sectors. We adopt an evolutionary perspective derived from the literature on evolutionary economic geography and economic complexity \citep{nelson_evolutionary_1982, boschma_evolutionary_2009, hidalgo_building_2009}. In doing so, we reframe the discussion on formal firms, casting them as sites of complex productive activities requiring teams of workers with diverse skills. We note that we do not directly measure the skill content or team composition of formal firms, leaving this to future work, but propose this mechanism as a modelling assumption and infer skill diversity indirectly via the geographical distribution of industrial activities using methods from \citet{hidalgo_building_2009}. 

The rest of this paper is organised as follows. Section 2 summarises the main strands of the economic literature related with this paper and provides background on Colombia, the country selected to test the hypothesis. Section 3 describes the main data sources and computation of variables, including our explanatory variable. Section 4 presents features derived from the data that are consistent with the hypothesis and the main econometric model. Section 5 concludes.

\section{Literature and background}

This paper is closely related to four literature strands. First, the literature on urbanisation economies, which emphasises the diversity of cities as the source of innovation and economic growth \citep{Jacobs1969, beaudry_whos_2009}. Second, the theoretical and empirical literature of evolutionary economic geography and economic complexity which sees development as a localised path dependent combinatorial process \citep{boschma_evolutionary_2009, hidalgo_building_2009}. Third, the literature on measuring industrial complexity and skills \citep{hidalgo_product_2007, Neffke2011regions}. And fourth, the economic development literature on the relation between formal and informal activities \citep{fields_rural-urban_1975, de_soto_other_1989, ulyssea_regulation_2010}. 

\subsection{Economic literature on cities}\label{sec_cities}

Our focus is on cities rather than countries. The size and growth of cities has been central research topics in urban economics, centred around the balance between agglomeration economies and congestion costs \citep{duranton2004micro, gabaix2004evolution, brinkman2016congestion}. The economic literature on cities has shown that a key reason why bigger cities generate more employment is the reduction in search friction costs, which allows better matches between workers and jobs \citep{mortensen_chapter_1999, moretti2010local, glaeser_triumph_2011}. However, skill diversity, and not just skill level or abundance, is central because significant agglomeration economies in production may result from the interaction between workers with different skills \citep{glaeser_learning_1999, Puga2010, glaeser_complementarity_2010} and because skill diversity is a precondition for industrial diversification \citep{harrison_specialization_1996, combes2000economic, beaudry_whos_2009}. Therefore, we emphasise the importance of skill diversity and not just the size of local labour markets and lower search friction costs.

We view formal firms as organised firms that combine complementary skills in order to produce sophisticated goods and services. Such firms thrive in cities, where large populations with diverse skills reside. While we do not directly show empirical support for this claim here, a number of studies support this stance. For example, using US data, \citet{bettencourt2014professional} find an increase in occupational diversity and complexity as cities grow larger. Using firm level data from Brazil, \citet{tian2021division} show that larger cities are host to firms with higher occupational diversity. 

We argue that the ability of a city to increase its formal employment rate relies on the ability of formal firms to access relevant specialised skills in related sectors. The tools and concepts of evolutionary economics and economic complexity, focusing on the evolutionary processes by which places combine existing and new --but proximate-- capabilities to diversify into new economic activities, provide both the conceptual framework and tools needed to understand and model the dynamics of formal employment creation.

\subsection{Evolutionary economics and economic complexity}\label{sec_evoecon}

Our hypothesis relies on two theoretical foundations that have wide empirical support from the literature on evolutionary economics and economic complexity. First, more complex industries demand not only more tasks, but a combination of distinct tasks, each requiring specialised skills. This idea was central to Adam Smith's treatise on the roots of the industrial revolution, as manufacturing firms emerged in the 1760s. Smith forecast that a division of labour, whereby workers were trained to have specialised skills, would result in higher productivity. Recently, a number of authors have documented increasing levels of individual specialisation \citep{jones2009burden} and investigated the relationship between worker productivity and skill complementarity in teams \citep{neffke_skill_2013, neffke2019value}. 

The second well-supported theoretical foundation is that firms evolve by tinkering with skills, because all feasible technologies and products are not known in advance but are discovered by recombining skills \citep[see][Chapter 11]{beinhocker_origin_2006}. The idea that industrial development relies on a combinatorial process has emerged in various forms, most notably from evolutionary economics. A related field, evolutionary economic geography, concerned with the conditions under which places develop, was an early adopter of these ideas. Pioneered by \citet{nelson_evolutionary_1982}, evolutionary economic geography argues that regions re-combine existing skills and capabilities in order to move into new economic activities which are similar to those already present \citep[see][]{boschma_evolutionary_2009}. A related literature focuses on the capabilities of nations and the emergence of complex exporting sectors \citep{hidalgo_product_2007, HausmannHidalgo2011}, and a third branch looks at the re-combination of technologies in the emergence of new innovations \citep{fleming2001technology}. 

A result of these evolutionary dynamics is a path dependent process in which regions and cities re-combine existing capabilities --and import new capabilities-- such that industries and technologies emerge and expand \citep{hidalgo_product_2007, Neffke2011regions}. A growing number of studies investigate path dependence in the structural transformation of regions (see \citet{zhu2019evolutionary} for a review). In particular, relatedness between economic activities \citep{hidalgo2018principle} fosters new production processes (from both mature and nascent firms), which naturally promotes the creation of new jobs, by exploiting synergies between skills and fostering the combination of complementary ideas \citep{delgado_clusters_2014, feldman1999innovation, Jacobs1969}. Many studies focus on firm and industry entry and growth. For example, \citet{neffke_skill_2013} using Swedish data, show that new firm survival and success depends on the proximity of the new firm’s industry to the local skill base, and \citet{diodato_why_2018} show that US cities depend increasingly on access to skills in related industries for employment growth.

We propose a framework in which employment in formal firms is characterised as an organised complex economic activity requiring the combination of complementary skills (as found in larger cities). In contrast, we see informal firms or independent workers as those who operate largely in low complexity activities in which the skills of different types of workers are not combined. This study aims to shed light on the mechanisms behind the creation of formal employment in complex sectors primarily through the growth of existing formal firms or the creation of new formal firms.\footnote{Formal employment growth in a city can come from four sources: inactive individuals transitioning into the labour force, unemployed individuals transitioning to employment, employed workers transitioning from informal to formal firms, and formal workers working more weeks per year (see Appendix 5 for a decomposition).} From our perspective, skill tinkering is more active and results in discovering more sophisticated products when firms have access to a larger and more diverse pool of skills in their catchment areas. Thus, the growth of formal employment in the form of complex industries is not random, but rather a path dependent process. 

While there have been a number of studies that deploy this type of approach to study industry and firm entry at the region or city level in developed country settings, there have been more limited applications in developing countries, as revealed by a recent review of the literature by \citet{zhu2019evolutionary}. Two studies that focus on Latin America are close to ours in their approach and methodology. First, \citet{hausmann_implied_2021} showed that the potential for employment growth in an industry is related to the basket of existing industries in Chilean municipalities. Second, \citet{jara2018role} focused on pioneer firms in Brazil (i.e., new firms operating in an industry that is new for the region) and found that the survival of these firms is dependent on hiring workers from closely related industries.

\subsection{Measurement of industrial complexity and skill relatedness}\label{sec_indcomplskrel} 

Operationalising these concepts is not trivial. Skills learned on the job and embedded in workers are both highly valuable, and difficult to measure. In the empirical literature on skills, a worker’s skill is usually equated to the worker’s educational attainment, and unobserved skill heterogeneity is dealt with by including controls for worker traits that are related to skills \citep{bacolod_elements_2010}. The use of educational attainment as a proxy for skills is at odds with abundant evidence on the inadequacy of education systems, especially in developing countries, to provide the skills required by formal firms \citep{ryan_school--work_2001, bassi_disconnected_2012, busso_learning_2017}, and on the importance of both cognitive and socio-emotional skills --as opposed to years of schooling per se-- on labour market outcomes \citep{glewwe_schools_2002, hanushek_schooling_2009}. 

We are interested in industry-specific skills mostly learned on the job, and thus adopt the ``industry complexity'' measure proposed by \citet{hidalgo_building_2009} and the ``skill relatedness'' measure developed by \citep{neffke_skill_2013, neffke2018agents}. Both studies take a ''phenotypic approach'', inferring the underlying genotype (e.g., skill proximity between industries) by observing the prevalence and combination of phenotypes (e.g., worker transitions). 

Industry complexity, an application of the well-established (export) product complexity of \citet{hidalgo_building_2009} to industry employment data, is an indirect measure of the complementary productive capabilities that an industry requires to operate. Industry complexity is computed based on an algorithm that takes as an input the geographic distribution of employment across industries and cities. The key idea is that industries that require many distinct and complementary capabilities will be located in places that produce many other things, including rare things. Hence, the algorithm infers the complexity of an industry from where it locates, and what else is made in that place.\footnote{For more details on algorithm behind the computation of industry complexity, see Section \ref{sec_measurement} and Appendix 4.} 

It is worth mentioning that the development of ``complexity indices'' remains an active topic of research, and several alternatives to Hidalgo and Hausmann’s measures have been studied and proposed \citep[see, e.g.,][]{tacchella_new_2012, cristelli_measuring_2013, mariani_measuring_2015, servedio_new_2018, mealy_interpreting_2019, brummitt_machine-learned_2020,rsoc2021estimating}. In particular, there have been recent efforts to more directly measure the type and diversity of capabilities needed for industries. Using detailed US data on tasks, \citet{turco2020knowledge} find that complex industries are most intensive in STEM knowledge and high skill tasks. They develop a new metric based on these STEM tasks that is both highly correlated with industry complexity and predictive of GDP per capita growth for US cities. However, detailed data on the task content of occupations and industries is rarely available, even for highly developed economies. We expect complex firms to host a wide array of skills and capabilities and thus exhibit high wage diversity. As a step towards validation of our industrial complexity metric from \citet{hidalgo_building_2009}, we show that firms in more complex industries have higher wage diversity (Theil entropy) at a firm level. 

Skill relatedness \citep{neffke_skill_2013, neffke2018agents} is a measure of the similarity of the skills used by any pair of industries, and is constructed via the number of worker transitions between them. It is one of a family of relatedness measures that aim to capture skill or knowledge proximity between sectors using a variety of approaches including co-production of goods in plants \citep{Neffke2011regions}, occupational similarity \citep{farjoun1994beyond, chang1996evolutionary} and co-presence on patent applications \citep{jaffe1989characterizing}. Skill relatedness is particularly suited to capture the proximity between industries as those industries exchanging workers are highly likely to require similar skills. Skill relatedness has several advantages over other approaches to measure the similarity of skills in that it uses very granular data and does not rely on extraneous classification systems, which may introduce noise \citep{diodato_why_2018}. Here we deploy this approach to measure the skill proximity of a city's industrial basket to those industries not already present.

\subsection{Economic development literature}\label{sec_econdev}

The early literature on development economics had a central geographical component. Authors such as \citet{lewis_economic_1954}, \citet{harris_migration_1970} and \citet{fields_rural-urban_1975} emphasised the dual nature of labour markets in developing countries. Within this literature, intense competition between workers for few urban formal jobs creates a large stock of unemployed workers who must resort to informal productive activities in urban areas. This early strand of literature took into account the relationship between rural and urban potential workers, but it did not recognise the diversity of production activities in urban areas and how they complement each other.

A second stream of economic development studies attributes the lack of formal jobs in developing countries to the way governments interfere in labour markets. It is argued that minimum wages, labour taxes, and other government interventions are important causes of informality because governments in developing countries lack enforcement capabilities \citep{de_soto_other_1989, rauch_modelling_1991, kugler_labor_2009, kugler_payroll_2017, levy_under-rewarded_2018, kanbur2017informality}. This argument has been extended and tested, for example, at the regional level in Brazil \citep{almeida2009enforcement, jonasson2012government} and Colombia \citep{fernandez2017impact, morales2017assessing}. The problem with this argument is that policies set at the national level, such as minimum wages in many developing countries, are not sufficient to explain the large variation in local employment formality rates. 

A third, and more recent, stream of development literature stresses the importance of frictions. Models of labour search and matching see informal employment as the result of labour market frictions between workers that are heterogeneous in their productivity and their needs of social protection. Firms may also be heterogeneous in their labour demand patterns and their visibility to government authorities. For both, labour regulations may generate benefits but imply compliance costs. Formal workers are protected by job security legislation and have access to the benefits of social security, but at the cost of paying income taxes and social security contributions. Likewise, benefits to firms include access to public goods, like the legal system and the police, while costs include taxes, regulations and bureaucratic requirements \citep{albrecht_effects_2009, bosch_comparative_2010, ulyssea_regulation_2010, meghir_wages_2015, narita2020}.

\section{Data and metrics}

\subsection{Data and basic variables}

Our main data source is the administrative dataset PILA (the Integrated Report of Social Security Contributions) managed by the Colombian Ministry of Health. It contains information on sole proprietors as well as employees in formal firms and organisations, by municipality and industry for 2008-2016, that contribute to social security. All types of sectors, including goods and services, are covered. We use data aggregated at four-digit industry codes, using ISIC (revision 3.0) with 345 industries.\footnote{We exclude from the analysis oil, mining, public administration and domestic service sectors because their size and employment growth are determined by factors largely unrelated to the mechanism we propose.}  

We define the formal occupation rate, $f_{c,t}$, as the ratio between $F_{c,t}$ --the number of full-year equivalent workers by private firms with two or more employees that report to PILA-- and $P_{c,t}$, the population 15 years or older in city $c$ in year $t$: 
\begin{equation}
    f_{c,t} = \frac{F_{c,t}}{P_{c,t}}
\end{equation}

Our unit of analysis is cities, defined as local labour markets, which can comprise more than one municipality, based on commuting patterns. To establish which municipalities constitute a multi-municipality city, or metropolitan area, we use the methodology proposed by \citet{Duranton2015delineating}, which he applied to Colombia. It consists of adding iteratively a municipality to a metropolitan area if there is a share of workers, above a given threshold, that commute from the municipality to the metropolitan area. We use a 10\% threshold, which in the case of Colombia for 2008 generates 19 metropolitan areas that consist of two or more municipalities (comprising a total of 115 municipalities). Similar to the standards of the US Office of Management and Budget (OMB) for metropolitan area delineations, we add to these 19 metropolitan areas another 43 individual municipalities that have populations above 50,000 inhabitants, for a total of 62 cities.

\subsection{Measuring industry complexity and proximity when skills are unobservable}\label{sec_measurement}

We adopt the methodology of \citet{hidalgo_building_2009} to measure industry complexity, an approach based on the distribution of industries across locations. Industry presence rests on a definition of Revealed Comparative Advantage (RCA) from \citet{balassa1965trade}, which we operationalise for our data on formal employment. In any given year, the RCA of an industry in a city is simply the share of formal employment in an industry in the city (relative to total formal employment in the city) divided by the share of formal employment in the industry (relative to total formal employment):
\begin{equation}
RCA_{c,i}=\frac{F_{c,i}/\sum_i F_{c,i}}{\sum_c F_{c,i}/\sum_c \sum_i F_{c,i}}
\end{equation}
where subscript $c$ indicates cities and subscript $i$ corresponds to industries. The RCA matrix is transformed to a binary matrix depending on whether a particular value is larger than 1 or not: $M_{c,i}=1$ if $RCA_{c,i}>1$ and otherwise $M_{c,i}=0$. Matrix $M_{c,i}$ indicates the industries that are more concentrated in a city relative to the national share. From this matrix we can compute the diversity of each city, that is, the count of the number of industries with $RCA_{c,i}>1$ for a given city $c$. We can also compute the ubiquity of each industry as the count of the number of cities with $RCA_{c,i}>1$ for a given industry $i$. The economic complexity index developed by \citet{hidalgo_building_2009} is derived iteratively from this matrix: an industry is complex if it is produced in few places (i.e., the ubiquity is low), but those places produce many things (i.e., they are diverse). Similarly, a city is complex if it produces many things (i.e., it is diverse), including things produced in few places (i.e., they have low ubiquity). Full details of this algorithm are provided in Appendix 4. 

In order to construct our key predictive variable, complexity potential, which quantifies the proximity of the current industrial base of a city to new complex industries, we require a measure of skill proximity between each pair of industries. As proposed by \citet{neffke_skill_2013}, labour flows (job switches) between two industries reveals their skill similarity. A high intensity of switches (relative to what would occur at random) can be measured as the ratio between the number of observed job switches between the industries and the expected job switches if they were random (i.e., using a network configuration model). Formally, if $\phi_{i,j}$ is the number of job switches between industry $i$ and industry $j$ (between year $t$ and year $t+1$), skill proximity can be computed as a matrix with entries:
\begin{equation}
    SP_{i,j} = \frac{\phi_{i,j}/\sum_{j}\phi_{i,j}}{\sum_{i}\phi_{i,j}/{\sum_{i,j}\phi_{i,j}}}
\end{equation}
Since this matrix is asymmetric, it is made symmetric by averaging with its transpose, and re-scaling the values so that they range from -1 to 1. This re-scaled matrix is stored as a new matrix $E$. Positive values of $E$ indicate that the switches between the two industries are larger than expected and are therefore taken as a measure of skill relatedness.\footnote{To be precise, if $(SR_{i,j} + SR_{j,i})/2 > 1$, or the average of the bi-directional switches between industries $i$ and $j$ is more frequent than expected, then $E_{i,j}>0$.} 

For each ``missing'' industry in a city (that is, industries in city $c$ for which $RCA_{c,i} < 1$), we can estimate the composite skill proximity to the set of industries that are ``present'' in the city (that is, industries with $RCA_{c,i} > 1$):
\begin{equation}
    dens_{c,i} = \frac{\sum_{j \in U_c} E_{i,j}}{\sum_{j} E_{i,j}}, \forall i \in R_c
\end{equation}
where $U_c$ denotes the set of industries present in city $c$ (RCA > 1) and $R_c$ denotes the set of ``missing'' industries (RCA < 1). This proximity measure $dens_{c,i}$ is known as ``density'' in the complexity literature \citep{hidalgo_product_2007, hausmann_implied_2021}.

Finally, our measure of complexity potential - the skill-proximity of a city to potential new complex industries - is computed as a weighted average of the complexity of the set of missing industries in a city, where the weighting corresponds to the density metric introduced above:
\begin{equation}
    CP_c = \frac{1}{|R_c|} \sum_{i \in R_c} dens_{c,i} * CI_i
\end{equation}
where $CI_i \in [0,1]$ is the normalised complexity of industry $i$ from \citet{hidalgo_building_2009} adapted to employment data as outlined in Appendix 4.
 
\subsection{Descriptive statistics}

Table \ref{table_1_2023} shows the industry complexity ranking of Colombian sectors (ISIC 2-digit codes or ‘divisions’). The complexity index for 2-digit ISIC codes is calculated as a weighted average of the complexity index at 4-digit ISIC codes, where the weights correspond to the size of employment. We observe that, in terms of manufacturing, plastics, clothing, chemical products, machinery, leather and textiles products top the ranking. In terms of services, informatics, financial services and real estate are the most complex sectors. 

\begin{table}[t!]
    \centering
    \includegraphics[width=0.80\linewidth]{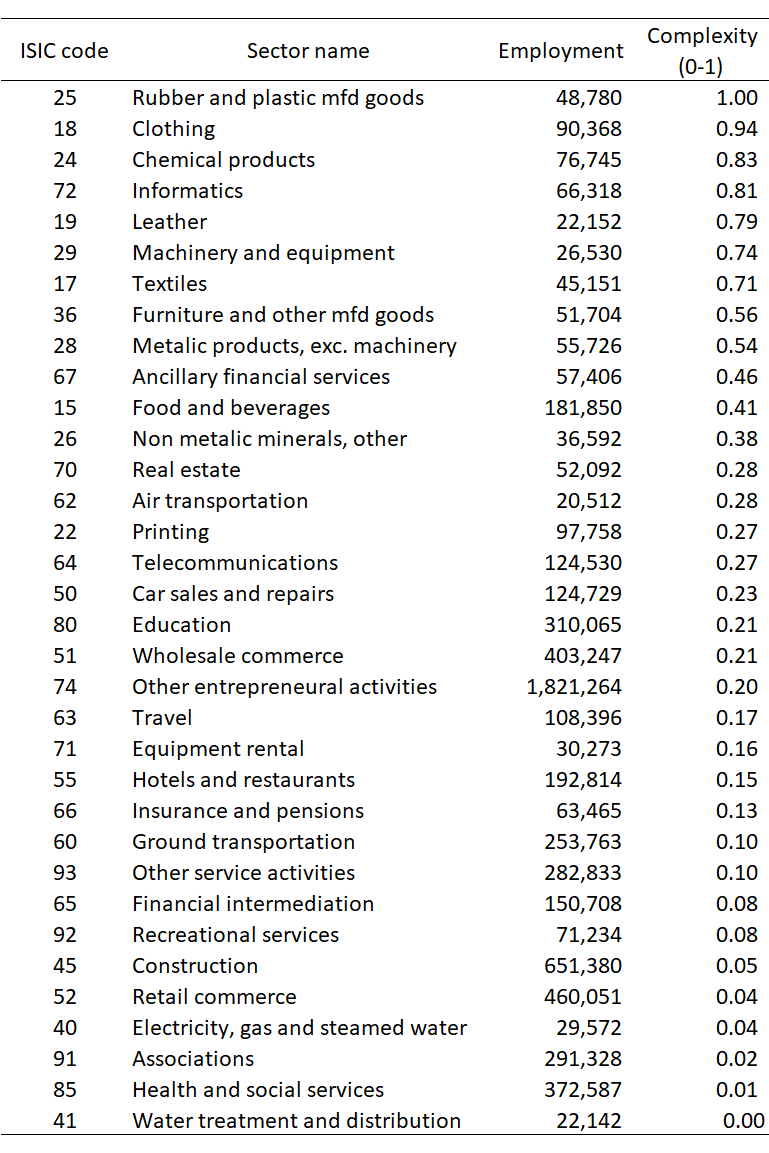}
\caption{Industry ranking of ISIC two digit codes (divisions) by mean normalized complexity (CI). Unipersonal firms, oil, mining, public administration and domestic service sectors are excluded (ISIC 10, 11, 75 and 95, respectively). Complexity is a weighted average of the industry complexities at the four-digit level, with weights given by formal employment.}
\label{table_1_2023}
\end{table}

Following \citet{hidalgo_building_2009}, complex industries are those located in few cities which produce a wide range of goods and services, a combination of low ubiquity and large diversity. For example, textiles, clothing and leather products are complex industries within the context of the Colombian economy because they are highly concentrated in the country's second largest metropolitan area and serve a dynamic export market. 

Table \ref{table_2_2023} presents summary statistics for the sample used in the econometric analysis. There are in total 496 city-year observations, which are the product of 62 cities and 8 years of data (2009 - 2016). The main goal of the econometric analysis is to test the relationship between the change in the formal employment rate in a city and its complexity potential.   

\begin{table}[t!]
    \centering
    \includegraphics[width=0.80\linewidth]{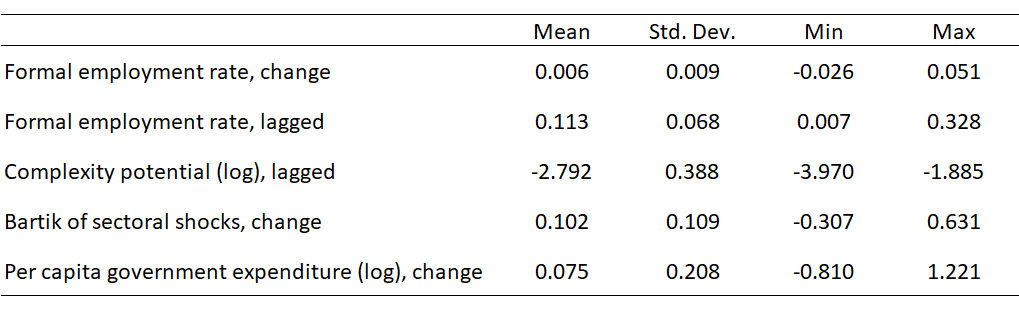}
\caption{Summary statistics. Number of observations: 496 (each observation is a city in a year). Unipersonal firms, and oil, mining, public administration and domestic service sectors are excluded.}  
\label{table_2_2023}
\end{table}

\section{Results}

\subsection{Formal employment, industry complexity and city size}

Our measure of industry complexity is a proxy for the degree of diversity and sophistication of the skills needed in each industry which are not observable \citep{hidalgo_building_2009}. This could potentially be measured (at least to some extent) via detailed task data, available in the US via ONET \citep{turco2020knowledge}, but unfortunately, high resolution data on tasks for all occupations is unavailable for most countries, and especially in developing countries such as Colombia. In order to validate our approach, we investigate the relationship between industry complexity and wage diversity. If the most complex industries coordinate teams of workers with diverse skills, then we would expect these to exhibit the highest wage diversity. Using firm level data, for firms with over 50 employees in 2014 in the PILA dataset, we compute the Theil coefficient of wage entropy for each firm. We find that, even after controlling for firm size, industry complexity is highly correlated with wage entropy (Table \ref{table_3_2023}). This also holds after controlling for a range of human-resource management related variables such as the average wage, the average workers' age, the standard deviation of workers' age, the personnel retention coefficient, the average wage increase, the standard deviation of wage increases, the share of women and the gender wage gap. 

\begin{table}[t!]
    \centering
    \includegraphics[width=0.80\linewidth]{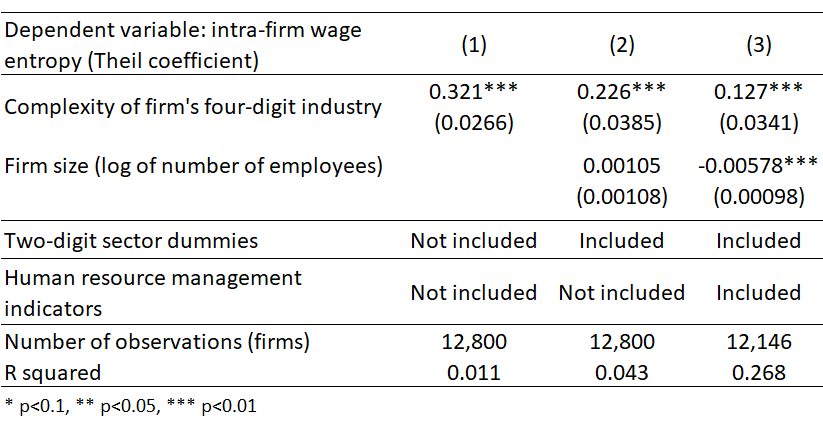}
\caption{Correlation between wage entropy and industrial complexity. Each observation is a firm with at least 50 employees. Oil, mining, public administration and domestic service sectors are excluded. Human resource management indicators are: average wage, average workers' age, standard deviation of workers' age, personnel retention coefficient, average wage increase, standard deviation of wage increases, share of women and gender wage gap. All data are for 2014 (personnel retention, wage increases and standard deviation of wage increases are computed with respect to 2013).}
\label{table_3_2023}
\end{table}

Workers in more complex industries also receive higher wages. Theoretical models with production functions characterised by skill complementarity predict higher wages in more complex industries \citep{kremer_o-ring_1993}. Empirical evidence for this prediction can be found in \citet{pekkarinen2002complexity} and \citet{rsoc2021estimating}. Table \ref{table_4_2023} provides simple econometric evidence for Colombia. Each observation corresponds to a firm-city-year, and the dependent variable is the log of the average real wage. The correlation between industry complexity and the average real wage within the industry is highly significant, even after the inclusion of city dummies, year dummies and firm's size (as measured by the effective number of employees).

\begin{table}[t!]
    \centering
    \includegraphics[width=0.80\linewidth]{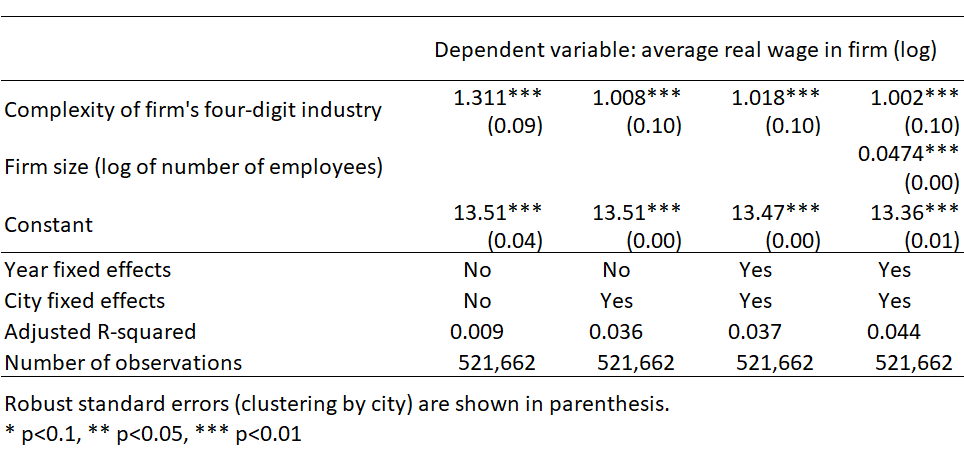}
\caption{Regressions of firm wages on industry complexity. Each observation is a firm in a city in a year. Unipersonal firms, oil, mining, public administration and domestic service sectors are excluded.}
\label{table_4_2023}
\end{table}

\begin{figure}[t!]
    \centering
  \includegraphics[width=0.49\linewidth]{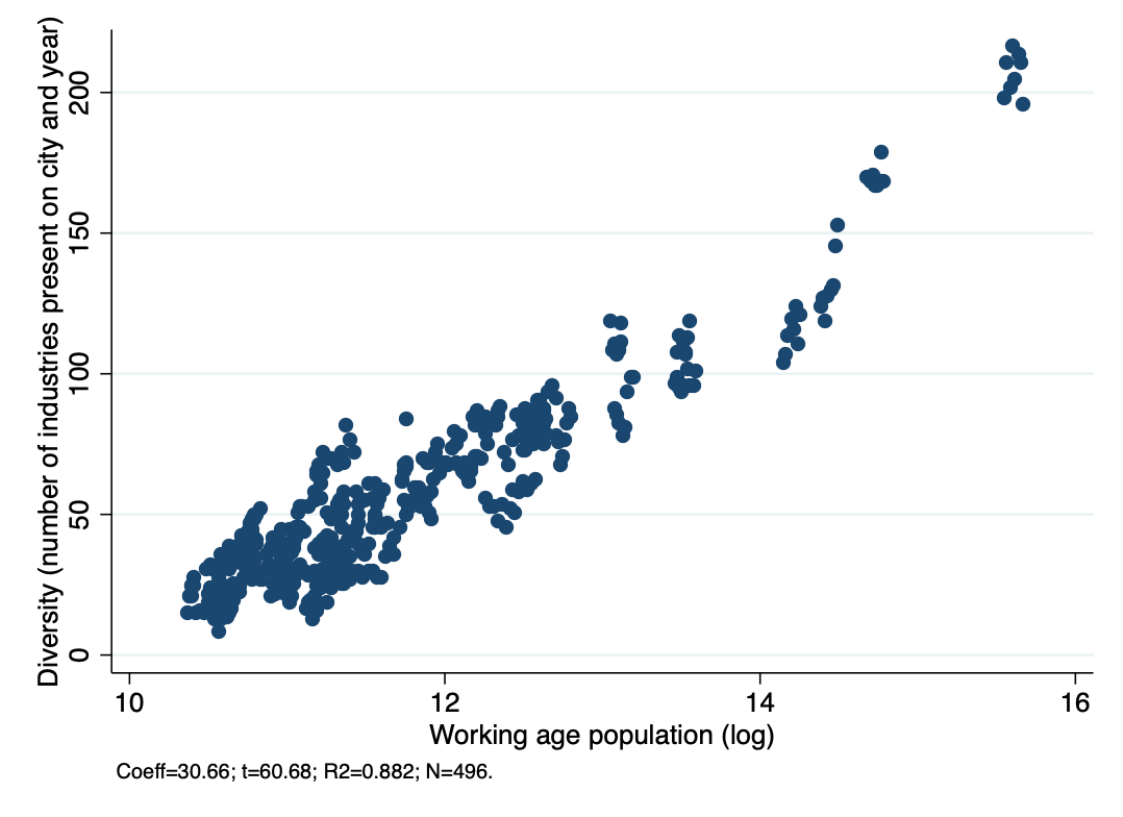}
  \includegraphics[width=0.49\linewidth]{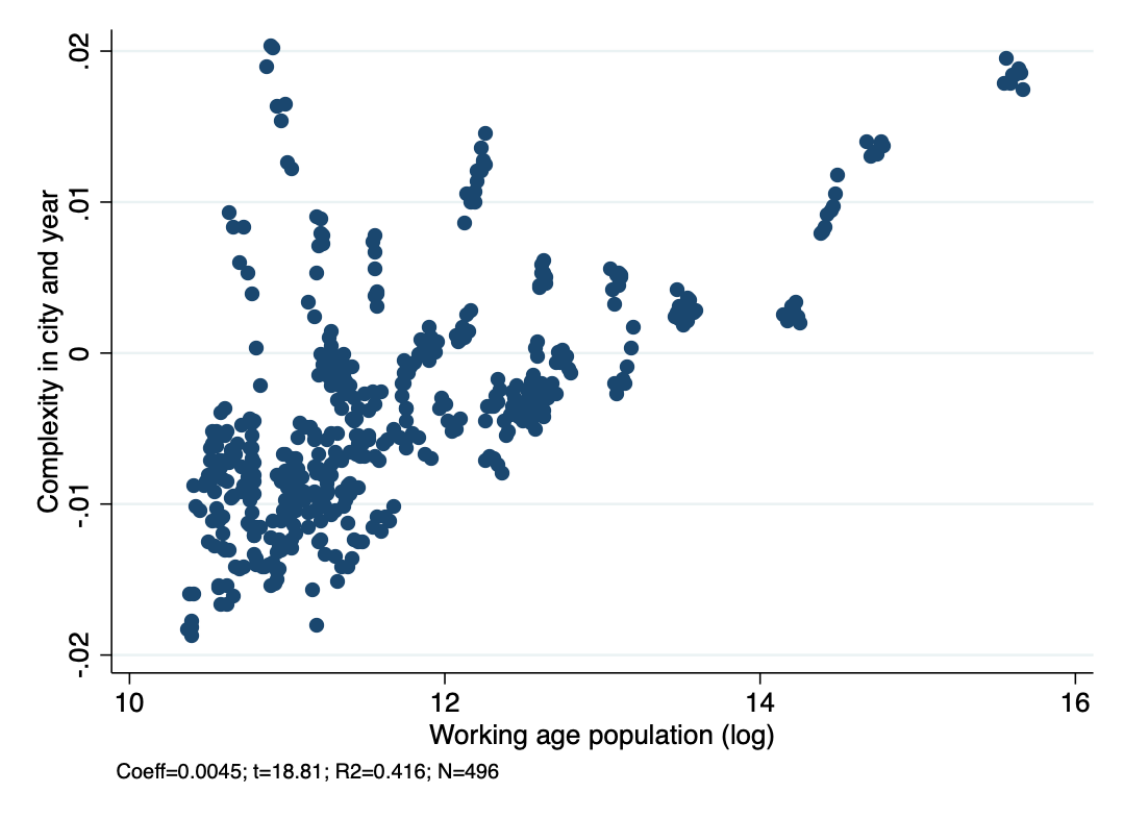}
     \caption{Left panel shows that cities with larger working age populations have a higher number of industries with $RCA>1$, our measure of diversity. Each point is a city-year. Likewise, right panel shows that larger cities have more complex industries (the vertical axis is the average complexity of the industries with $RCA>1$). }
\label{fig_3_2022}
\end{figure}

Next, we present some salient features of formal employment in cities in Colombia that are consistent with our hypothesis, which will be formally tested in the next section. Our starting point is the relationship between population size and diverse and complex ecosystems of economic activities, which has been noted elsewhere \citep{quigley_urban_1998, henrich2004demography, kline2010population, bettencourt2014professional, youn_scaling_2016, muthukrishna_innovation_2016, gomez-lievano_explaining_2017, rsoc2021estimating}. We show in Figure \ref{fig_3_2022} that larger cities have more diverse and complex industries. The correlation between the number of industries that are present in a city (e.g., have RCA>1) and the city's working age population (in logs) is 0.88, as shown in Figure \ref{fig_3_2022} (left panel), where each point represents a city-year observation. While Bogot\'{a}, the largest city, has between 195 and 216 industries (depending on the year of observation), the smallest city (Fundaci\'{o}n) has between 14 and 27 industries. Figure \ref{fig_3_2022} (right panel) also shows that larger cities have, on average, more complex industries with correlation of 0.42. The vertical axis is the average complexity of the industries present in the city and the horizontal axis is the working age population (in logs). 

\begin{figure}[t!]
    \centering
    \includegraphics[width=0.95\textwidth]{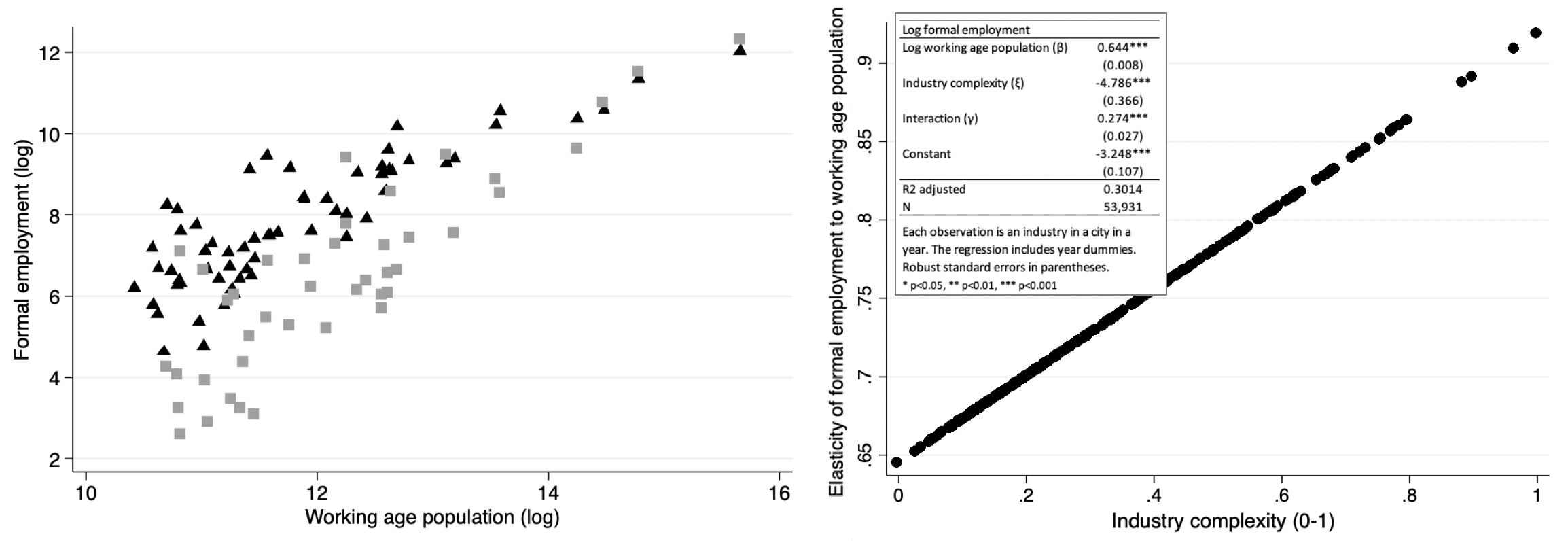}
   \caption{The left panel shows formal employment versus working age population by city in 2015 for low complexity industries (industries in bottom complexity decile, triangles) and high complexity industries (industries in top complexity decile, squares). The steeper slope in the latter case indicates that employment in more complex industries increases more rapidly with city size. The right panel shows more systematically the elasticity of industry-city employment to city size as a function of industry complexity as given by the regression in Equation \ref{eqn_elast1}. The inset in the figure presents the table of results from this regression, which uses annual data between 2008 and 2015 (year fixed effects are included in the regression). ***p=0.001}
\label{fig_4_2022}
\end{figure}

Closely related to the previous feature is a second one: employment in more complex industries is disproportionately higher in larger cities. In order to show this, we have classified in deciles of complexity all the industries that are present in at least one city (independently of their size). Figure \ref{fig_4_2022} (left panel) shows the relationship between formal employment and the working age population in cities for high complexity industries (top decile) and low complexity industries (bottom decile). We observe a steeper slope, or increased elasticity, in the former case, suggesting that the response of employment to city size (in a cross-section) is more pronounced in the more complex industries. This feature is consistent with previous results from \citet{gomez-lievano_explaining_2017}, \citet{balland2020complex} and \citet{davis_comparative_2020}.

In order to systematically examine the changing distribution of labour by industry complexity, we compute the elasticity of industry-city employment to city size as follows. We run a regression for formal employment using variation across cities $c$ and industries $i$:
\begin{equation}
\label{eqn_elast1}
log(F_{c,i})=\alpha + \beta log(P_c)+\xi q_i + \gamma log(P_c)CI_i+\epsilon 
\end{equation}
where $P_c$ is the working age population of city $c$, and $CI_i$ denotes the complexity of industry $i$. We then compute the elasticity coefficients (i.e., taking the derivative with respect to $log(P_c)$):
\begin{equation}
\label{eqn_elast2}
\frac{\partial log(F_{c,i})}{\partial log(P_c)}=\beta + \gamma CI_i   
\end{equation}

The right panel of Figure \ref{fig_4_2022} shows the results of the regression based on Equation \ref{eqn_elast1} and plots the elasticity coefficient as a function of industry complexity given by Equation \ref{eqn_elast2}. We observe that formal employment in more complex industries rises more rapidly as the working age population increases, suggesting that large cities incubate a larger proportion of complex industries than smaller cities. This does not imply that more complex industries generate, by themselves, more employment than less complex ones (since total employment in complex industries across all cities might be less than total employment in simpler industries).

\begin{figure}[t!]
    \centering
    \includegraphics[width=0.7\linewidth]{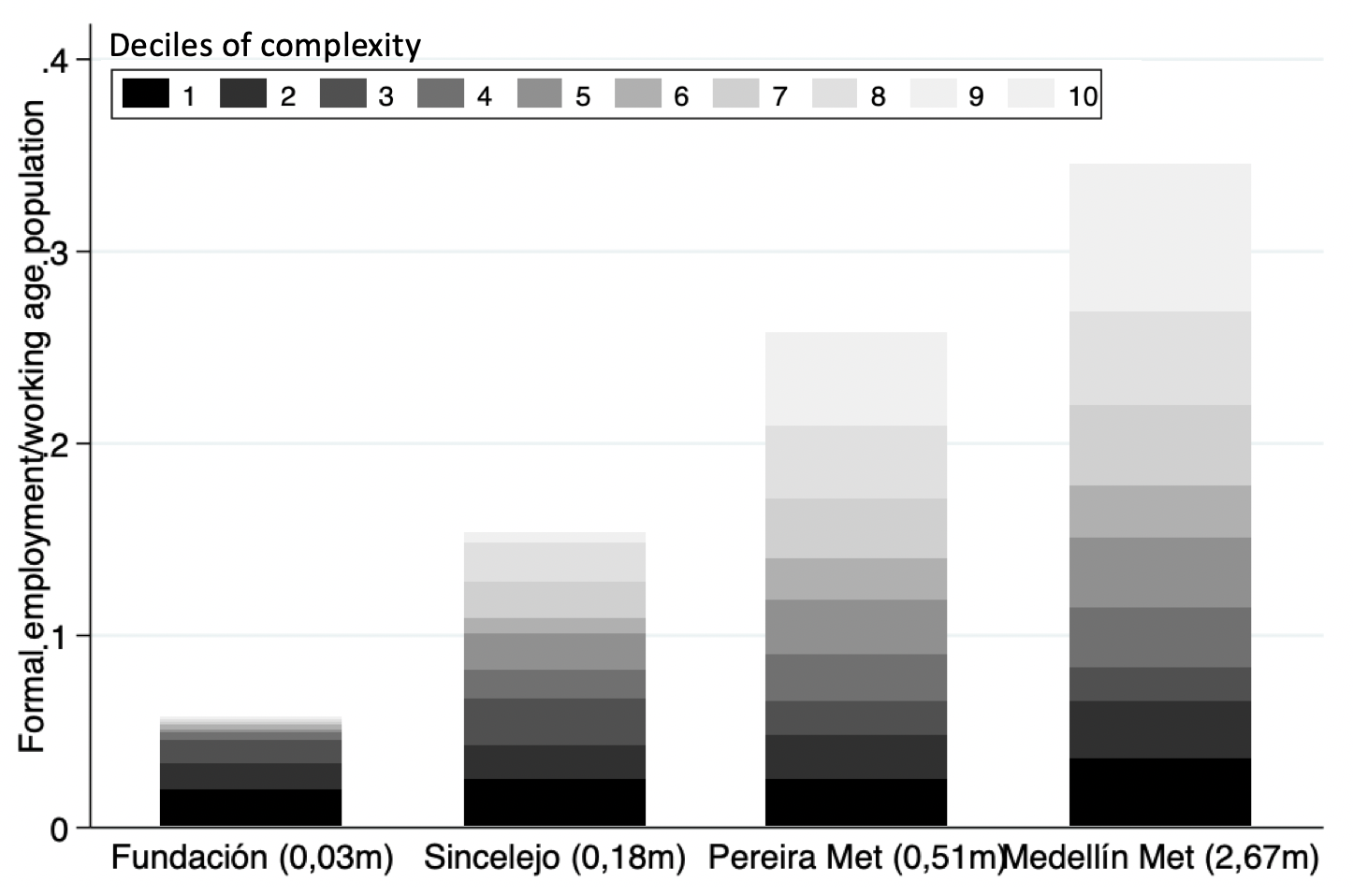}
   \caption{``Geological layers'' of formal employment rates in four cities of different sizes (working age population shown in parenthesis) in 2015. Low complexity sectors have similar relative shares independent of city size, while high complexity sectors constitute a larger share in larger cities.}
\label{fig_5_2022}
\end{figure}

Finally, Figure \ref{fig_5_2022} provides a graphic illustration of the resulting ``geological layers'' of employment due to the fact that more complex industries, though smaller than less complex ones, are disproportionately larger in more populated cities. Four cities of different sizes have been selected (Fundación, Sincelejo, Pereira and Medellín), and for each the formality rates by complexity decile are presented (recall that the formality rate of each industry in a city is defined as the ratio between its number of full-year equivalent employees and the city's working age population). Larger cities have higher total formality rates due to the larger size of their high complexity industries. Notice that the lowest deciles are approximately the same relative size in the four cities.

\subsection{Econometric results}

\begin{figure}[t!]
    \centering
    \includegraphics[width=0.47\linewidth]{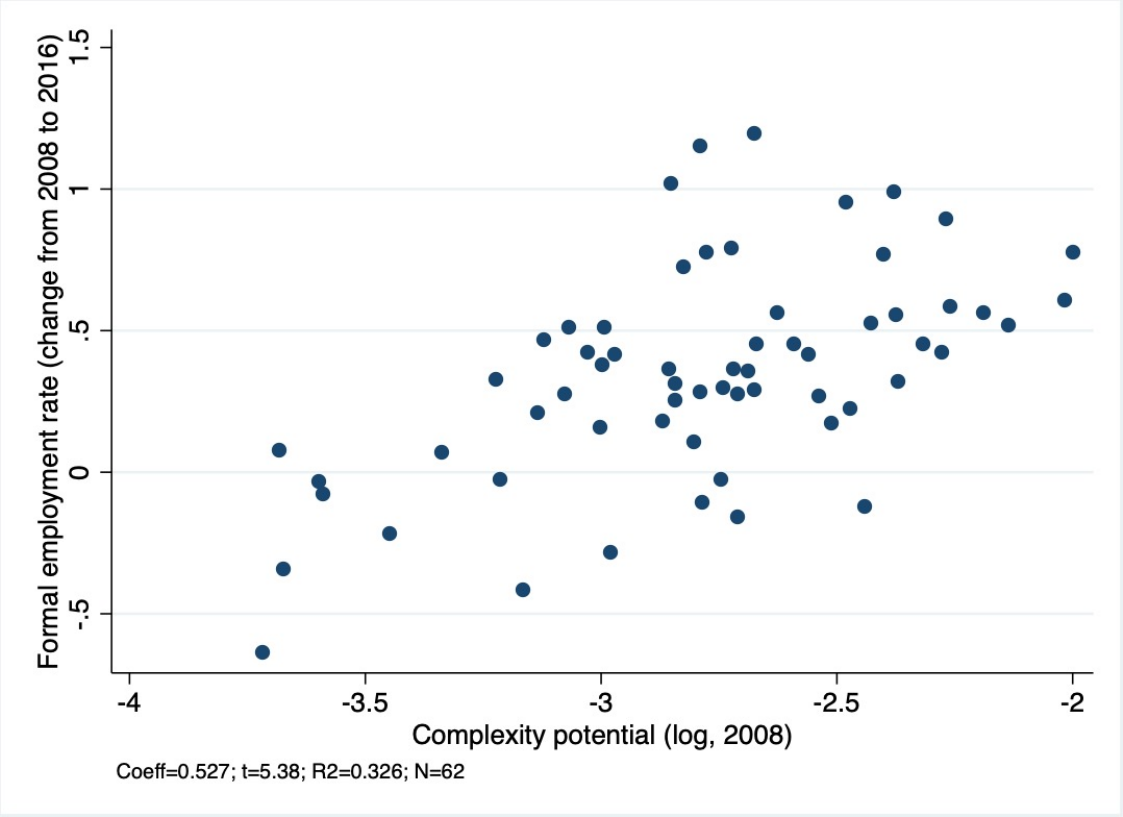}
   \caption{Correlation between complexity potential and subsequent formal employment growth. The higher the initial complexity potential of a city in 2008, the larger the increase, until 2016, of its formal employment rates.}
\label{fig_6_2022}
\end{figure}

Consistent with the trends documented in the previous section, the main hypothesis of this paper is that the ability of firms within a city to create formal employment depends on the proximity of the current formal industrial base to new skills required by complex industries not already present in the city. Figure \ref{fig_6_2022} shows that the higher the initial complexity potential of a city in 2008, the larger the increase, between 2008 and 2016, of its formal employment rate. 

Table \ref{table_5_2023} presents the main econometric results. The sample consists of 496 city-year observations with data for 62 Colombian cities across 8 years (2009 - 2016). The dependent variable for all columns is the annual change in the formal employment rate. Column 1 shows that cities in which a larger share of the workforce are formal employees experience a larger change in the formal employment rate. Column 2 shows the main correlation between a city's complexity potential and the change in its formal employment rate over the next year. We can interpret this as showing that cities that have skills (in industries) closer to complex industries not present in the city are more likely to experience a larger change in their formal employment rate. Thus, a city's lagged complexity potential is positively correlated with the change in its formal employment rate.\footnote{Table \ref{table_5_2023} also reports the Akaike information criterion (AIC) and the Bayesian information criterion (BIC) for all models. The model reported in column 2 has the minimum AIC and BIC amongst the models with complexity potential as a regressor.}

\begin{table}[t!]
    \centering
    \includegraphics[width=0.8\linewidth]{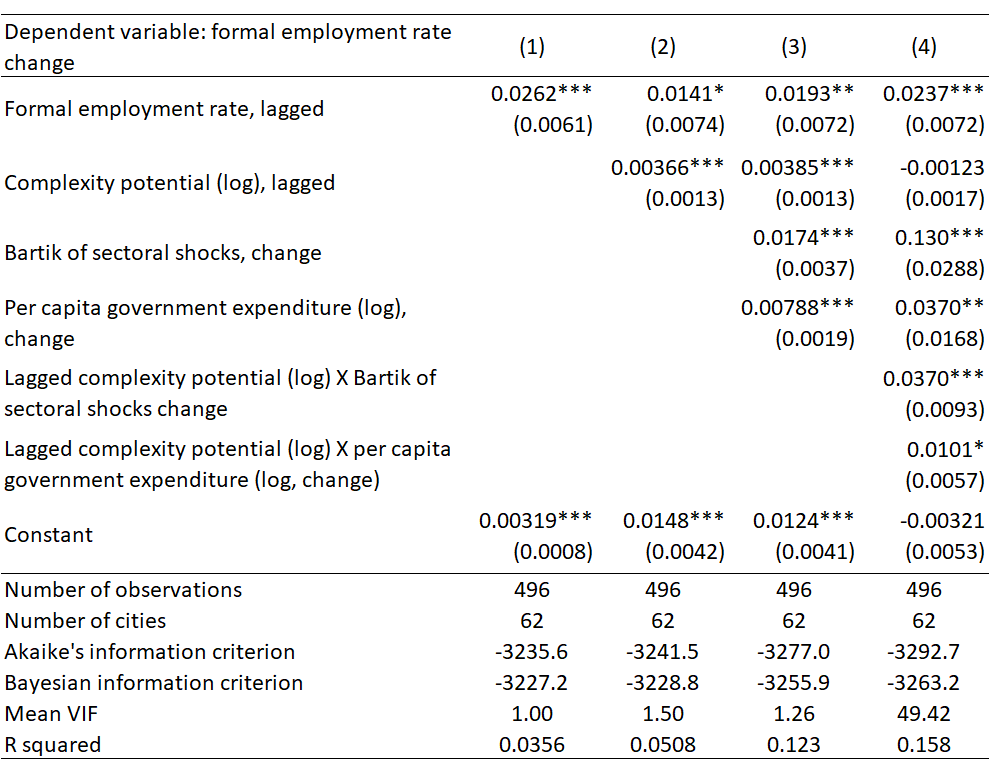}
    \caption{Correlation between initial complexity potential and formal employment change. Each observation is a city. Formal employment does not include oil, mining, public administration or domestic service sectors.}
\label{table_5_2023}
\end{table}

Column 3 in Table \ref{table_5_2023} explores the correlation between complexity potential and formal employment when taking into account supply and demand shocks experienced by the city's local economy. Nation-wide supply shocks are captured through a Bartik-style instrument\footnote{Appendix 3 provides details on the methods of calculation of this variable, following \citet{bartik_who_1991}.}, whereas demand shocks are measured as the change in per capita government expenditure. Both demand and supply shocks are positively correlated with the change in formal employment rate. The positive correlation between complexity potential and the change in the formal employment rate holds when including both proxies for supply and demand shocks. Column 4 shows that the positive correlation between complexity potential and formal employment growth does not hold when it is interacted with both proxies for supply and demand shocks. In this case, a larger change in formal employment rate occurs in cities with a larger interaction term between complexity potential and the proxy for supply shocks. However, the variance inflation factor reported at the bottom of Table \ref{table_5_2023} indicates that there is strong collinearity among the regressors included in column 4 (VIF = 49.42). Therefore, our preferred specification in Table \ref{table_5_2023} are the results reported in column 3.\footnote{Appendix 7 presents the results of an alternative dynamic estimator in which the dependent variable is the formal employment rate as a function of its lag and lags of complexity potential. These econometric estimations are not part of the main results because they failed to pass the Sargan test.}

To test the robustness of the relationship between formal employment creation and complexity potential, we consider two variables that may influence the ability of firms to create formal employment which have been previously tested for some developing countries by \citet{almeida2009enforcement, jonasson2012government, quiroga2021education}: institutional quality and higher education quality. Better institutional quality may facilitate formal employment creation through several channels: by improving the provision of public goods, such as infrastructure, security and justice, that raise the productivity of formal firms, by reducing macroeconomic risks, by reducing the administrative costs of creating formal firms, and by effectively enforcing the labour code including the collection of social security contributions. Similarly, by increasing productivity, the quality of higher education may facilitate formal employment creation. We use measures of institutional quality and higher education quality computed by Universidad del Rosario and \citet{de2011indice} for 26 of the 33 Colombian departments (states). We have attributed the data for the corresponding department to 55 cities. Our measure of institutional quality was originally computed as a simple average of five indices: administrative performance, fiscal management, government transparency, red tape for private business, and security and justice. Our measure of higher education quality is the original score computed by \citet{de2011indice}, following the methodology of the Global Competitiveness Report by \cite{schwab2013global}.

\begin{table}[t!]
    \centering
    \includegraphics[width=0.8\linewidth]{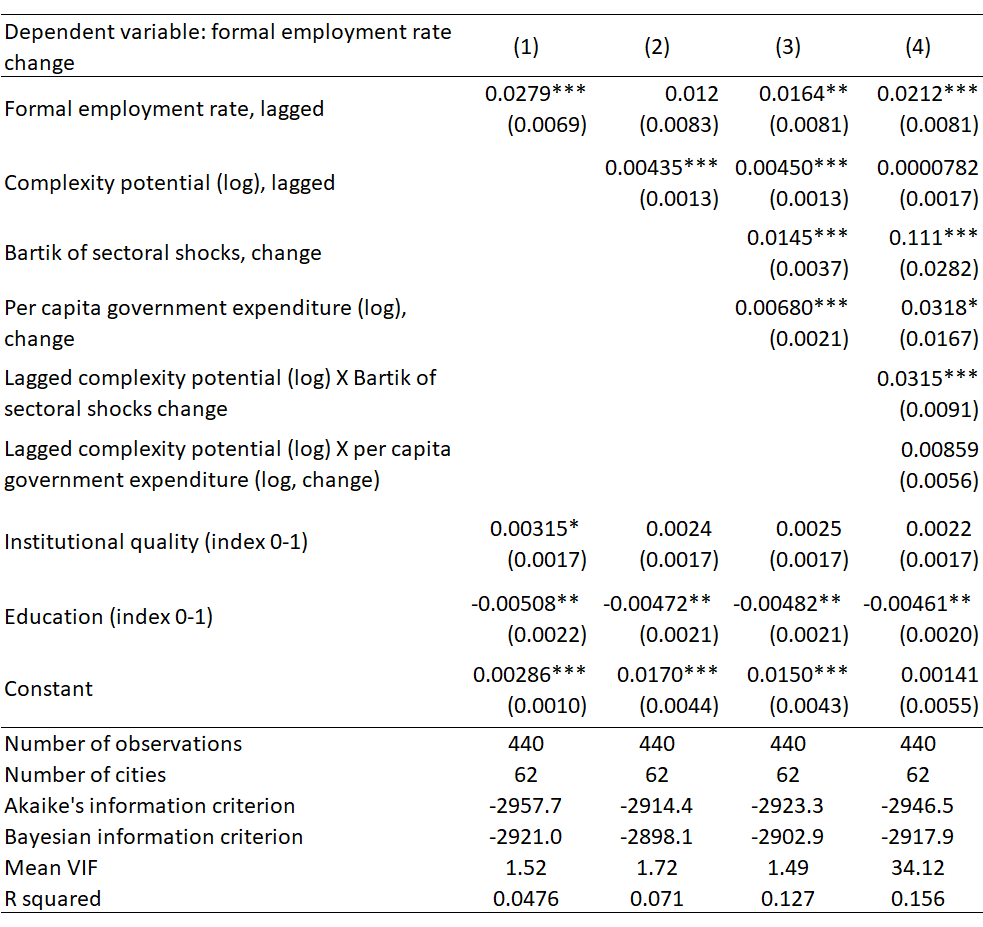}
    \caption{Robustness test for the correlation between initial complexity potential and formal employment change. Each observation is a city. Formal employment does not include oil, mining, public administration or domestic service sectors.}
\label{table_6_2023}
\end{table}

Table \ref{table_6_2023}, which has the same structure as Table \ref{table_5_2023}, presents the results of including our measures of institutional quality and higher education quality into the analysis. The positive and significant correlation between complexity potential and the change in the formal employment rate holds after controlling for these two variables (columns 2 and 3).

Overall, our estimates support the hypothesis that the capacity of cities to create formal employment is constrained by the local availability of the mix of skills needed to develop industries not already present.

\section{Discussion}

We have proposed an evolutionary mechanism to understand why larger cities in developing countries are able to generate higher rates of formal employment. Specifically, we show that the growth of formal employment is dependent on the potential of a city to enter new complex industries by building on existing skills.

\subsection{Relationship with the international literature}

This paper contributes to a number of literatures and current debates. Our work first and foremost sits within a broad literature in economic geography and economic complexity which takes an evolutionary perspective on regional industrial transitions, seeing growth as the outcome of a path dependent process whereby regions build on existing knowhow embedded in the workforce \citep{nelson_evolutionary_1982, boschma_evolutionary_2009, hidalgo_product_2007}. We extend this thinking to a developing country context in which a relatively small formal sector sits alongside a massive informal sector. In doing so, we bring into sharp focus the role of local characteristics, and in particular skills and labour market conditions, in the growth of the formal sector which has been largely overlooked in previous literature. Hence, our findings represent a synergy between economic complexity theory, the industrial policy research community, and employment growth, as discussed by \citet{hidalgo2021economic} and \citet{ferraz2021linking}.

Our focus on cities, and their characteristics is rooted in both the urban economics literature and the urban scaling literature. These fields see large cities as ‘cauldrons’ in which a huge diversity of specialised workers interact \citep{Romer1986, Lucas1988, Romer1990, Kremer1993population}. This diversity and these interactions lead to disproportionately high levels of innovation and creativity, and are the key drivers of economic growth in modern service led economies. Consistent with what we would expect, although not previously documented in the literature, we find that the share of formal employment similarly rises disproportionately with city size. But this employment is not added uniformly: larger cities ‘add’ formal employment in ‘geological layers’ of complexity such that the share of formal employment in complex sectors increases with city size. These findings open a new avenue through which to investigate the formal economy through an urban lens.  

We also contribute to the literature on the informal economy. Firstly, we develop a conceptual framework in which formal firms are organised activities characterised by teamwork done by workers with distinct specialised skills, while informal firms are more likely to consist of individuals with more general skills. Hence, while we construct and empirically test a model for the growth of the formal sector, we also contribute to a better understanding of the characteristics of the informal sector. Secondly, moving away from a traditional focus on the determinants of firm transitions from informal to formal, we instead bring to the fore the role of local labour market conditions and industrial concentrations in the growth of formal employment. 

More generally, we contribute to a resurgent emphasis on the urban dimension in the economic development literature, yet with a novel focus. Although central to the early literature on economic development \citep{lewis_economic_1954, harris_migration_1970}, the link between urbanisation patterns and formal employment growth is almost entirely absent in contemporary literature. Reciprocally, labour informality is absent from the burgeoning urban economic literature focused on developing countries (see the reviews by \citealp{brueckner_chapter_2015, desmet_geography_2015, chauvin_what_2017}), although housing informality does attract some attention (see \citealp{smolka_housing_2011} for a review). 

\subsection{Limitations}

We note that there are some important limitations of our study. First, additional research is needed to uncover the causal relationship between the skill-proximity of the current industrial base to new complex industries and the degree of formalisation in local labour markets. We recognise that the sources of variation in the complexity potential of Colombian cities are not exogenous. The dynamic models are a good approach to tackle the endogeneity concerns in the cross-sectional analyses. Future research should take into account the recommendation by \citealt{baumsnowferrerira2015} for causal inference in urban and regional economics.

As a second limitation, our results hinge on measuring correctly and accurately the relatedness between industries and their complexities. Reducing the measurement error in these quantities, in principle, should only strengthen the statistical significance of our results, but it is worth emphasising that the measures we used are proxies, and future work should be devoted to improving the estimation of the underlying fundamental quantities. Related to this, as a third limitation, future work is needed to empirically validate our conjecture that formal firms are indeed host to complex teams, while informal firms are characterised by low complexity activities. 

As a fourth limitation, Colombia is a country that shares many similarities with other developing economies, but it is also a unique country, including its legacy of internal armed conflicts. Thus, more data and research are needed that cover other geographies. Finally, as a fifth limitation, we studied a time window (from 2008 to 2016) that is short relative to some economic processes that have time-scales that act over many more decades.\footnote{\citet{mcnerney2021bridging}, however, show that the short-term process of related diversification is informative of the long-term process of economic development.} 

\subsection{Policy recommendations}

While we elaborate on the wider policy implications in Appendix 8, we briefly summarise the main points here. Firstly, industrial policy should focus on policies which support firms in accessing and developing local skills to move into new sectors. Beyond traditional skills support such as on-the-job training and mobility schemes, this could involve better transport links to enlarge the spatial scope of the effective labour market, or importing workers with specialist skills via internal or international inwards migration or FDI. 

Secondly, policies aiming to grow the formal sector should move away from a sole focus on the transition of small firms from informality to formality. This means less focus on indicators of red tape such as the Doing Business ranking, and national level labour policies such as minimum wages, labour taxes and social security legislation. While these may - at least temporarily - increase the size of the formal sector by encouraging informal firms to transition, this narrow focus neglects the main source of formal employment growth at the other end of the scale as formal firms expand their activities in complex sectors.  

\begin{footnotesize}
\bibliographystyle{agsm}
\bibliography{Cities.bib}
\end{footnotesize}



\section*{Supplementary material}

# Appendixes for "Industrial complexity and the evolution of formal employment in developing cities"

## Appendix 1 - Tying together strands of literature

In this paper we add to the discussion on the mechanisms that underlie the growth of formal employment rate at a city level. Distinct from previous work, we do not focus on the transition of informal firms to formality, but rather we focus on job creation at the other end of the spectrum in complex sectors. In doing so, we tie together ideas and results from a wide spectrum of literatures, introduced above. Here we bring together those ideas into a summary of our key contributions.

The idea that larger cities are home to more complex sectors (Balland et al. 2020) and firms with more diverse skills (Tian 2021) is not in doubt. Our contribution is to link city size and industrial complexity to higher rates of formal employment in cities in developing countries. We propose that formal activities are inherently organised activities while informal activities tend to be family-run and less complex (with some key exceptions found in complex illegal activities such as drug trafficking). Inspired by issues around defining informality via a purely legal lens, this definition finds support in official documents from India, in which the government refers to formal firms as organised firms (NCEUS 2007). Similarly, it is consistent with the established notion of firms as contractors across teams of tacit skills (Coase 1993, Grant 1996), and burgeoning work on the nature of firms as coordinating teams of complementary workers (Jones 2009, Neffke 2019).

We are concerned with the mechanisms that underlie the growth of formal employment, via the growth of existing formal firms and the entry of new ones, in cities. It is well established via a large literature emerging from evolutionary economic geography that cities enter new sectors and experience employment growth in sectors that are related to what they already know how to do (Hidalgo et al. 2007, Neffke et al. 2011). Closest to our work, Jara-Figueroa et al. (2018) find that the survival and growth of pioneer firms in Brazilian cities depends crucially on hiring workers with experience in related activities. We harness these ideas to show that the rate of formal employment creation in developing cities is dependent on the local presence of related skills in workforce. To measure relatedness between sectors, we rely on an existing method which consists of tracking worker job switches between industries pioneered by Neffke & Henning (2013).

Finally, a key challenge for this work is the issue of how to measure or quantify the diversity of skills present in the local labour market in a developing country context for which detailed data is scarce. While we do not measure skills directly here, instead using established methods to infer industrial complexity via geographical patterns of production. Consistent with our theory, Tian (2021) use occupational data from Brazilian firms to show that larger cities, which are home to more complex industries, are also home to firms with more diverse occupations. We show below, using Colombian data, that more complex industries are home to firms with higher wages and higher levels of wage diversity (which would be expected if they have more diverse skills). We deduce from these facts that industrial complexity can be used as a proxy for skill-diversity, albeit an imperfect one.

## Appendix 2 - Colombia

Colombia is a fitting case study, not only because of data availability, but due to several features of its economic geography, its labour market and institutions. With 76 percent of its 47.1 million inhabitants residing in urban areas (in 2013, as estimated by DANE, the National Statistical

1

Office), Colombia has 62 cities with at least 50,000 inhabitants (see definition of cities below). Geographic diversity and differentiated patterns of colonisation have resulted in geographically-varied levels of government capabilities, public goods provision and prosperity still visible today (Acemoglu et al. 2015). However, major increases in public social expenditures since the Constitution of 1991 have significantly reduced gaps in access to education and health across cities.

Overall, we believe that the patterns we detect here are representative of a general trend, particularly in fast growing Latin American cities, but there are a variety of factors present in the Colombian case which may be important and not be found elsewhere, such as tax changes and the mining boom, during our period of analysis. Our findings may also be affected by the fact that Colombia experienced an internal armed conflict that lasted more than fifty years between the mid- 1980s and 2018, and caused 7.5 million people, most of them with low levels of education, to migrate from the areas affected by the conflict to mostly urban areas (Lora 2019).

**Appendix 3 – Data sources**

**Data for Colombia**

Our *main data source* is the administrative data PILA (the Integrated Report of Social Security Contributions), managed by the Colombian Ministry of Health. It contains information on formal employment and wages by firms, municipalities and industries for 2008-2015. All types of sectors, including goods and services, are included (note that we use the terms 'sector' and 'industry' interchangeably). We use data aggregated at four-digit industry codes, using ISIC (revision 3.0) with 390 industries, and 62 cities.

*Cities* are defined using the methodology by Duranton (2015). Applying his algorithm to population and commuting data by municipality from the 2005 Census by DANE (National Statistics Office) we obtain 19 urban areas that consist of two or more municipalities, comprising a total of 115 municipalities. Similar to the standards of the US Office of Management and Budget (OMB) for metropolitan area delineations, we add to these 19 urban areas another 43 individual municipalities with urban populations above 50,000 inhabitants in 2008 according to DANE, for a total of 62 cities.

*Dependent variables in regressions*

- The dependent variable in the regressions in Table 1 is *wages*, expressed as logs of the ratio between firms' average wages and the minimum wage. Computations are made by industry-year with a sample of 100,000 firms randomly selected from the universe of all the firms observed in the whole period (from the social security data mentioned above).
- The dependent variable in the regressions of Tables 2, 3 and 4 is the annual change in the *formality rate* as defined in the text (taking two-year averages o the corresponding initial and final rates), and computed from the social security system data mentioned above and working age population data (see below).

*Explanatory variables in regressions*

- The regressions in Tables 1 use the measure of *industry complexity* explained in Appendix 1, computed from the social security data mentioned above.
- All the explanatory variables in the regressions of Tables 2, 3 and 4 are two-year averages.

2

- The main explanatory variable in the regressions of said Tables 2, 3 and 4 is *complexity potential*, computed as explained in the text from the social security system data mentioned above.
- *Working age population* (for 2008-2015), defined as population 15 or older, is calculated from population data by age groups and municipality estimated by DANE. This variable is also used to compute the formality rate by city.
- *Oil producing city*, a binary variable, which takes the value of one if the city has more than one oil well in production per thousand inhabitants. Oil well data refers to 2014, as reported by Ecopetrol (the Colombian hydrocarbon company) for their own internal records.
- *Government spending shock* is the change between t0 and t1 (within 2008- 2015) in total government spending (in 2008 prices) per working-age person. It is computed from municipality-level government spending data compiled by CEDE.
- *Sectoral demand shocks,* $sds_c$, is a Bartik-style measure (see Bartik, 1991) that summarizes for each city the mix of nationwide sectoral demand shocks facing the city. It is computed as

$$sds_c = FOR_c \sum_i \frac{emp_{c,i}(t0)}{emp_c(t0)} g_{i,c}$$

where $g_{i,c} = log[emp_i(t1)] - log[emp_i(t0)]$ is growth of employment of industry *i* excluding employment in industry *i* in city *c*. In other words, $emp_i = \sum_{j \in J} emp_{i,j}$ with set *J* containing all cities except city *c*.

- *Quality of institutions, infrastructure quality and higher education quality* for 2011 (the earliest year available) are taken from Consejo Privado de Competitividad and Universidad del Rosario. This data was collected for 26 of the 33 Colombian departments following the methodology of the Global Competitiveness Report by Schwab and Sala-i-Martín (2013). We have attributed the data for the corresponding department to 54 cities (the departments of the remaining seven cities were not covered).

**Data for Mexico, Brazil and USA for Figures 1 and 2**

Data on formal employment for Mexico is publicly available (by municipality) from the Mexican Atlas of Economic Complexity produced by the Centre for International Development at Harvard (accessible at www.complejidad.datos.gob.mx). Population by age groups is also publicly available for Mexico from INEGI. We use the official definition of 'Metropolitan Zones' from Conapo, for which there are 59 zones.

Formal employment data for Brazil comes from RAIS (by municipality) via Dataviva (www.dataviva.info). The working age population data (by municipality/age group) is from census data 2010. A list of the 82 most populous cities based on the population of the municipality where the city is located was obtained from IBGE.

Formal employment data for the USA is provided (by metropolitan area) from the County Business Patterns for 2013. The working age population is from the US Census by county. We use the standard definition of Metropolitan Statistical Areas for the US from the US Office of Management and Budget, of which there are 388.

**Data for other Developing Countries for Figure 2**

The data on formality rates over a collection of developing countries is based on information from SEDLAC (CEDLAS and The World Bank) for countries in Latin America and the Caribbean, and International Labor Organization (ILO) for other countries. SEDLAC defines informal workers as salaried workers in firms of less than 5 employees, non-professional self-employed and unpaid workers. ILO defines informal workers as those who in their main or secondary jobs are own-account workers, employers and members of cooperatives or employed in their own informal sector enterprises. Working age populations (15 years or older) are taken from the same sources.

**Appendix 4 – Calculation method for industry complexity**

In this appendix we describe the methods for calculating the industry complexity variable introduced in section 3.2, which are adapted from Hidalgo and Hausmann (2009). The actual calculations used formal employment data of all industries producing either goods or services (ISIC-AC, Rev. 3, at 4 digits, using the national registry of the social security administration for Colombia, known as PILA). In the following equations, sub-index $c$ indicates cities and sub-index $j$ indicates industries. While we do not use a sub-index for years in order to simplify notation, all calculations are applied for each year separately (2008-2015).

*Calculation of Revealed Comparative Advantages*

We start with data for employment by industry, city and year, organized in matrix form:

$X_{cj}$ = number of employees in industry $j$ in city $c$

We include only industries with at least 50 formal employees in an average month, and only cities with at least 10 industries with 50 or more formal employees.

From this matrix, we compute aggregates, as follows:

$X_c = \sum_j X_{cj}$ (A2.1)

$X_j = \sum_c X_{cj}$ (A2.2)

$X = \sum_c \sum_j X_{cj}$ (A2.3)

These metrics are used to calculate the Revealed Comparative Advantage (RCA) for each city/industry combination:

$RCA_{cj} = \frac{X_{cj} / X_j}{X_c / X}$ (A2.4)

*Diversity and Ubiquity Calculations*

The RCA matrix is transformed into a binary matrix depending on whether a particular value is larger than 1 or not:

$$M_{cj} = \begin{cases} 1 & RCA_{cj} \geq 1 \\ 0 & RCA_{cj} < 1 \end{cases}$$

Matrix $M_{cj}$ indicates the industries that are relatively large in each city. From this matrix we compute the diversity indicator at the city level, and the ubiquity indicator at the industry level – that is, the count of the number of industries with relatively large employment for each city, and the count of the cities that have a given industry with a relatively high intensity:

$$k_{c,0} = \sum_j M_{cj} \quad k_{j,0} = \sum_c M_{cj}$$

*Industry Economic Complexity*

The complexity of an industry can be measured by its ubiquity weighted by the diversity of the localities that have revealed comparative advantage in such industry. Extending this exercise *ad infinitum*, correcting diversity with ubiquity and vice-versa with consecutive iterations, is called the *method of reflections*. It can be expressed as follows:

$$k_{c,0} = \sum_j M_{cj} \quad k_{j,0} = \sum_c M_{cj}$$

$$k_{c,1} = \frac{1}{k_{c,0}} \sum_j M_{cj} k_{j,0} \quad k_{j,1} = \frac{1}{k_{j,0}} \sum_c M_{cj} k_{c,0}$$

…

$$k_{c,n} = \frac{1}{k_{c,0}} \sum_j M_{cj} k_{j,n-1} \quad k_{j,n} = \frac{1}{k_{j,0}} \sum_c M_{cj} k_{c,n-1}$$

Substituting for $k_{c,n-1}$ on the right-hand side, we can obtain a closed expression for the industry complexity at iteration *n*:

$$k_{j,n} = \frac{1}{k_{j,0}} \sum_c M_{cj} \left( \frac{1}{k_{c,0}} \sum_{j'} M_{cj'} k_{j',n-2} \right)$$

$$= \sum_{j'} k_{j',n-2} \sum_{c} \frac{M_{cj'} M_{cj}}{k_{c,0} \, k_{j,0}}$$

$$= \sum_{j'} k_{j',n-2} \, \widetilde{M}^J_{j,j'}$$

where

$$\widetilde{M}^J_{j,j'} \equiv \sum_{c} \frac{M_{cj'} M_{cj}}{k_{c,0} \, k_{j,0}}$$

Using vector notation, the method can be written in a compact manner as:

$$\vec{k}_n = \widetilde{M}^J \times \vec{k}_{n-2}$$

When $n \to \infty$, $\vec{k}_n = \vec{k}_{n-2} = \vec{k}$, and we obtain the following expression:

$$\widetilde{M}^J \times \vec{k} = \vec{k}$$

We can obtain a solution for $\vec{k}$ (the vector of industry complexity values) by finding the eigenvector corresponding to the second largest eigenvalue of matrix $\widetilde{M}^J$.

### Appendix 5 – Sources of formal employment growth

The mechanism put forward in the paper requires the presence of untapped sources of labour supply so that formal employment may expand. While the economic literature on labour informality focuses on the movement of occupied workers between the formal and the informal segments of the economy, other sources of additional labour supply are plausible. Recall that the formal employment rate in city $c$ at year $t$ ($f_{c,t}$) is defined as the number of workers effectively employed by formal firms ($F_{c,t}$) divided by working age population ($P_{c,t}$):

$$f_{c,t} = \frac{F_{c,t}}{P_{c,t}} \qquad \text{(A1.1)}$$

The expression can be expanded in the following way:

$$f_{c,t} = \frac{F_{c,t}}{P_{c,t}} = \frac{L_{c,t}}{P_{c,t}} * \frac{O_{c,t}}{L_{c,t}} * \frac{N_{c,t}}{O_{c,t}} * \frac{F_{c,t}}{N_{c,t}} \qquad \text{(A1.2)}$$

where $L$ is labour force, $O$ is employed labour force and $N$ is the number of those who are employed by formal firms at least some fraction of the year. Therefore, the formal employment rate is the product of four components:

$$f_{c,t} = p_{c,t} * (1 - u_{c,t}) * v_{c,t} * i_{c,t} \qquad \text{(A1.3)}$$

where $p_{c,t}$ is the labour force participation rate (*L/P*), $u_{c,t}$ is the unemployment rate (*U/L=1-O/L*), $v_{c,t}$ is the formality rate of the occupied (*N/O*) and $i_{c,t}$ is the fraction of the year that the formally occupied dedicate to their employment (*F/N*). The former two fractions constitute the extensive margins of labour supply, and the two latter constitute the intensive margins. Using data for 23 cities for 2016 (from monthly household surveys by DANE, the official Department of Statistics), the corresponding average rates are (in the same order as in equation A1.3):

0.199 = 0.642 * 0.8779 * 0.5214 * 0.6739

The relevant concepts are defined as follows:

- *P* (working age population) are those 15 or older
- *L* (labour force) comprises the occupied (*O*) and the unemployed (*U*)
- *O* (occupied) are those 15 or older who worked at least an hour in the reference period (past week)
- *U* (unemployed) are those 15 or older who did not work at least an hour in the reference period but actively looked for a job in the past month
- *N* (formal workers) are the occupied (*O*) who are employees and contribute to social security (the self-employed are not considered employees)
- *F* (effective formal workers) are the formal workers (*N*) multiplied by *m*/12, where *m* is the number of months that they worked in the past year.

Formal employment in Colombia effectively absorbs less than a fifth of the full-time working potential of the working age population as the rates of participation, employment, formality, and dedication are all significantly less than 1. All these margins may operate simultaneously allowing organized employment to grow. Thus, taking logs and time differentiating yields the following decomposition of the sources of formal employment growth:

$$\Delta ln f_{c,t} = \Delta ln p_{c,t} + \Delta ln(1 - u_{c,t}) + \Delta ln v_{c,t} + \Delta ln i_{c,t} \qquad (A1.4)$$

With average annual data for the 23 cities between 2008 and 2016, the average contribution of each of the four sources to formal employment growth in Colombia is as follows (in the same order as in equation A1.4):

0.0683 = 0.0058 + 0.0021 + 0.0407 + 0.0198

As the decomposition indicates, higher formality rates (of the occupied) have been the largest, but not sole, labour supply source feeding the increase of formal employment in recent years. The decomposition indicates, also, that the unemployed are but a minor source of additional labour supply.

## Appendix 6 – Correlation matrix

| Variable number | Variable | (1) | (2) | (3) | (4) | (5) | (6) | (7) |
|---|---|---|---|---|---|---|---|---|
| (1) | Formal employment rate change | 1 | | | | | | |
| (2) | Formal employment rate lagged | 0.1887 | 1 | | | | | |
| (3) | Lagged complexity potential (log) | 0.2094 | 0.5757 | 1 | | | | |
| (4) | Bartik of sectoral shocks change | 0.1779 | -0.1171 | -0.1028 | 1 | | | |
| (5) | Per capita government expenditure (log, change) | 0.1615 | -0.1052 | -0.0505 | 0.0509 | 1 | | |
| (6) | Lagged complexity potential (log) X Bartik of sectoral shocks change | -0.128 | 0.1704 | 0.2169 | -0.9855 | -0.0469 | 1 | |
| (7) | Lagged complexity potential (log) X per capita government expenditure (log, change) | -0.1404 | 0.121 | 0.0996 | -0.0486 | -0.9924 | 0.0514 | 1 |

Number of observations: 496 (each observation is a city in a year). Unipersonal firms, and oil, mining, public administration and domestic service sectors are excluded.

## Appendix 7 – System dynamic panel-data estimations of formal employment rates

| Dependent variable: formal employment rate, 2009-2016 | (1) | (2) | (3) | (4) | (5) |
|---|---|---|---|---|---|
| Formal employment rate lagged | 1.003*** | 0.942*** | 0.966*** | 0.905*** | 0.934*** |
|  | (0.0143) | (0.0373) | (0.0152) | (0.0231) | (0.0249) |
| Complexity potential (log) | 0.00662** | 0.0138*** | 0.00679*** | 0.0152*** | 0.0135*** |
|   One lag | (0.0028) | (0.0047) | (0.0025) | (0.0042) | (0.0044) |
|   Two lags |  | 0.0181** |  | -0.00712 | -0.00543 |
|  |  | (0.0091) |  | (0.0064) | (0.0072) |
|   Three lags |  | -0.00236 |  | 0.0106 | 0.00499 |
|  |  | (0.0117) |  | (0.0066) | (0.0082) |
|   Four lags |  | -0.00728 |  | -0.00923 | -0.00928 |
|  |  | (0.0076) |  | (0.0050) | (0.0056) |
| Bartik of sectoral shocks |  |  | 0.0115*** | 0.0283*** | 0.0227*** |
|  |  |  | (0.0040) | (0.0062) | (0.0078) |
|   One lag |  |  |  |  | 0.00609 |
|  |  |  |  |  | -0.00735 |
|   Two lags |  |  |  |  | 0.00312 |
|  |  |  |  |  | -0.00903 |
| Per capita government expenditure (log) |  |  | 0.00275 | 0.00508*** | 0.00485 |
|  |  |  | (0.0017) | (0.0020) | (0.0030) |
|   One lag |  |  |  |  | -0.00453 |
|  |  |  |  |  | -0.00407 |
|   Two lags |  |  |  |  | 0.00328 |
|  |  |  |  |  | -0.00457 |
| Constant | 0.0258*** | 0.0763*** | -0.209** | -0.546*** | -0.646*** |
|  | (0.0090) | (0.0273) | (0.0824) | (0.1350) | (0.1980) |
| Year dummies included? | Yes | Yes | Yes | Yes | Yes |
| Number of observations | 434 | 310 | 434 | 310 | 310 |
| Number of cities | 62 | 62 | 62 | 62 | 62 |
| Number of instruments | 62 | 42 | 132 | 98 | 82 |
| Sargan test of overidentifying restrictions | 0 | 0 | 0 | 0 | 0 |

Each observation is a city in a year. Unipersonal firms, and oil, mining, public administration and domestic service sectors are excluded. All the regressors (other than the year dummies) are treated as endogenous variables for the estimation.

Standard errors (generalized method of moments) in parentheses. * $p<0.1$, ** $p<0.05$, *** $p<0.01$

## Appendix 8 – Policy Recommendations

From a policy perspective, our findings imply that in order to accelerate the process of formal employment creation in cities, the first step should be identifying the barriers that retard or inhibit firms from tinkering with skills to improve their chances of survival and growth, since this is the mechanism that, in our view, leads to the apparition of more complex industries. When possible, barriers should be removed and ways to accelerate the evolutionary process should be put in place at the city level. Therefore, industrial policies should be incremental, based on a mapping of the technological similarities between sectors (such as those presented in the Colombian Atlas of Economic Complexity), and aimed at facilitating the apparition of sectors relatively close to

those already present in the city. These policy implications are similar to those of evolutionary economic geography, which sees regional economic development as a localised gradual, path-dependent process (see, e.g., Boschma 2014 for similar arguments), and suggests, e.g., that innovation policies should aim at accelerating knowledge transfer at the regional level between technologically related sectors (Boschma 2009).

To facilitate the growth of formal employment via industrial transitions in cities, a key emphasis needs to be placed on developing and importing skills. Although it is tempting to equate education with skill development, a burgeoning literature suggests that skills learned on the job are critical (Ryan 2001, Bassi et al. 2012), and cannot be substituted by formal education or training (Glewwe 2002, Hanushek & Woessmann 2009). Hence, while lack of education is probably one of the factors behind the size of the informal sector, national- level education and training policies may easily miss the target of endowing the labour force with the diversity of skills required at the local level to facilitate the apparition of more complex industries. Beyond developing skills, firms also need to be able to access these skills in large congested urban areas. Hence, the location of firms and households in the city, and the means of transportation available to workers, are important factors. O'Clery et al. (2019) show that a commute time of at most 40-60 minutes is optimally associated with formal employment growth for Colombian cities, with a range of 45-120 minutes in the UK (Straulino et al. 2023). Furthermore, since the size of the local labour market matters for formal employment growth, policies aimed at expanding public transportation systems beyond municipal borders should be encouraged. This is often impeded by the lack of coordination between municipalities and the difficulty of creating metropolitan areas that effectively operate as such.

If key skills and capabilities are missing in the local labour force, inward migration might provide a solution. Attracting scarce skills from other cities or countries seldom is a policy priority, while labour in-migration is often a source of concern, especially if immigrants do not easily fit into the local culture. However, research from Bahar & Rapoport (2018) shows that international migration is associated with successful 'import' of skills leading to entry of new sectors. In this respect, national governments could subsidise the relocation of skills across cities, the reason being that reallocating workers to cities where their skills are scarce generates an externality through the local emergence of new industries. Since such workers do not capture in higher wages the externality, a subsidy is socially warranted. This does not imply that highly-educated workers would be the main beneficiaries, since technical skills are often more scarce than professional skills in middle-income countries. Similarly, attracting foreign professionals in scarce fields at the city-level should be a component of a well-designed policy of international migration. However, in developing countries, professional accreditation requirements are often insurmountable for foreigners and are usually administered by national-level entities without regard for city-level needs.

Finally, more generally, policies aiming to grow the formal sector should move away from a sole focus on the transition of small firms from informality to formality. In this vein, the focus on the formalisation of small businesses by easing registration requirements and related policies encouraged by the World Bank's Doing Business rankings may not deserve the attention it receives from policymakers, even though those policies may (temporarily) increase the number of those firms and workers classified as formal. Second, the importance attached to national-level policy instruments, such as minimum wages, labour taxes and social security legislation, often overriding any other possible interventions, should be questioned. Although policies that raise labour costs may partly be responsible for labour informality (Levy 2010, Kugler & Kugler 2009, Kugler et al. 2017, Levy 2018), they should not be seen as the main or the most important cause

of the problem. Further research is needed to understand how/if these policies affect the speed of formal employment creation (and not just the level of formal employment or the share of formal occupation in total employment).

.

\end{document}